\documentclass[12pt]{iopart}
\usepackage{iopams}

\newcommand{\Hamop}[0]{\hat{H}}
\newcommand{\om}[0]{\omega}
\newcommand{\inth}[0]{\int^\infty_0}
\newcommand{\intf}[0]{\int^\infty_{-\infty}}
\newcommand{\dom}[0]{{\rm d}\omega}
\newcommand{\dk}[0]{{\rm d}k}
\newcommand{\cop}[0]{\hat{c}}
\newcommand{\copa}[0]{\hat{c}^\dagger}
\newcommand{\aop}[0]{\hat{a}}
\newcommand{\aopa}[0]{\hat{a}^\dagger}

\newcommand{\bop}[0]{\hat{b}}
\newcommand{\bopa}[0]{\hat{b}^\dagger}
\newcommand{\bopv}[0]{\hat{\bi{b}}}
\newcommand{\bopva}[0]{\hat{\bi{b}}^\dagger}
\newcommand{\Dop}[0]{\hat{\bi{D}}}
\newcommand{\Bop}[0]{\hat{\bi{B}}}
\newcommand{\Eop}[0]{\hat{\bi{E}}}
\newcommand{\Hop}[0]{\hat{\bi{H}}}
\newcommand{\Pop}[0]{\hat{\bi{P}}}
\newcommand{\Dv}[0]{\bi{D}}

\newcommand{\dv}[0]{\bi{d}}
\newcommand{\Bv}[0]{\bi{B}}

\newcommand{\rv}[0]{\bi{r}}
\newcommand{\nabv}[0]{\boldsymbol{\nabla}}
\newcommand{\real}[0]{{\rm Re}}
\newcommand{\imag}[0]{{\rm Im}}
\newcommand{\drv}[0]{{\rm d}^3\bi{r}}

\begin{document}

\title[Effective photons]{Effective photons in weakly absorptive dielectric media and the Beer-Lambert-Bouguer law}

\author{A C Judge$^1$, J S Brownless$^1$, N A R Bhat$^2$, J E Sipe$^3$, M J Steel$^4$ and C Martijn de Sterke$^1$}

\address{1 Centre for Ultrahigh bandwidth Devices for Optical Systems (CUDOS), Institute of Photonics and Optical Science (IPOS),
School of Physics, The University of Sydney, NSW 2006, Australia}
\address{2 BMO Capital Markets, 100 King St. W., Toronto, Ontario M5S 2A1, Canada}
\address{3 Department of Physics and Institute for Optical Sciences, University of Toronto, Toronto, Ontario M5S 1A7, Canada}
\address{4 Centre for Ultrahigh bandwidth Devices for Optical Systems (CUDOS), MQ Photonics Research Centre, Department of Physics and Astronomy, Macquarie University, NSW 2109, Australia}
\ead{a.judge@physics.usyd.edu.au}
\begin{abstract}
We derive effective photon modes that facilitate an intuitive and convenient picture of photon dynamics in a structured Kramers-Kronig dielectric in the limit of weak absorption. Each mode is associated with a mode field distribution that includes the effects of both material and structural dispersion, and an effective line-width that determines the temporal decay rate of the photon. These results are then applied to obtain an expression for the Beer-Lambert-Bouguer law absorption coefficient for unidirectional propagation in structured media consisting of dispersive, weakly absorptive dielectric materials.
\end{abstract}

\maketitle

\section{Introduction
}
The quantum theory of linear macroscopic electrodynamics in dispersive and dissipative media is by now a mature field of research. Although a quantised theory of the vacuum electromagnetic field was an early result of quantum mechanics, a rigorous treatment accounting for the presence of a macroscopic medium has proven to be a more elusive goal. Although many authors have contributed to this development \cite{Lax:1971,Drummond:1990,Glauber:1991,Nelson:1994,Matloob:1995,Gruner:1996,Dung:1998,Dung:2003,Suttorp:2004,Bhat:2006,Suttorp:2007,Raabe:2007,Philbin:2010,Judge:2013}, formal interest in this problem is often traced to the work of Jauch and Watson \cite{Jauch:1948}, while a significant advance in developing a canonical approach was made by Huttner and Barnett \cite{Huttner:1992}. Among the applications to date of the fully developed theory are the calculation of Casimir forces \cite{Philbin:2011} and spontaneous emission rates \cite{Yao:2009} in the presence of exotic materials. The general approach taken in canonical treatments of the quantisation procedure \cite{Suttorp:2004,Bhat:2006,Philbin:2010} is to couple the vacuum electromagnetic field to a set of harmonic oscillators representing the medium, which are continuously distributed in frequency $\om$ and position $\rv$. Diagonalisation of the uncoupled field and medium Hamiltonians introduces characteristic bosonic excitations corresponding to each: photons in the case of the field and, following the terminology introduced by Hopfield \cite{Hopfield:1958}, polaritons in the case of the medium. The eigenoperators of the coupled system therefore correspond to photon-polaritons, or dressed photons (DPs). In this closed system, dissipation of the electromagnetic field then manifests as a transfer of energy from the photons to the medium, with the rate of dissipation determined by the coupling strength. In the DP picture the eigenoperators are associated with spatial mode field distributions. Thus, since the DP operators vary harmonically in time, decay of the field results from dephasing of an initial superposition of DP modes.

The concept of dressed operators in the treatment of dissipative systems is well developed in quantum optics \cite{Barnett:1997}. However, despite the simple dynamics of the DP operators, employing them in order to determine the macroscopic electromagnetic field evolution is not straightforward: a complicating factor is the highly over-complete, non-orthogonal nature of the DP mode field basis. This implies that a given initial field distribution generally coincides with a multitude of initial states of the system, and assumptions regarding the initial state of the medium must therefore be made in order to uniquely determine the system evolution. Furthermore, one may conceive of states where the energy of the system initially resides entirely within the medium and is subsequently transferred to the electromagnetic field as the dynamics unfold. Such initial states, though physically permissible, are of little practical interest. Since in many applications one does not have control over the microscopic state of the medium, it is desirable to identify methods that avoid physically irrelevant configurations of the system in a convenient fashion. 

Considering experimental quantum photonics, the optical materials employed may generally be considered to be weakly absorptive over the bandwidth of interest. Indeed, the analysis of spontaneous nonlinear processes in integrated platforms (one of the major programs in integrated quantum photonics) is typically built upon a linear treatment of the field-medium interaction in the dispersive but non-absorptive regime \cite{Yang:2008}. Importantly, the structure of such linear theories \cite{Drummond:1990,Bhat:2006} is identical to that of the vacuum case \cite{Berestetski:1984}; that is, the degrees of freedom of the medium need not ultimately be considered explicitly in order to construct and evolve the electromagnetic field.  The causal electromagnetic response
functions are enough.

In establishing a path to including absorption in these treatments of quantum photonic processes, here we identify the coherent dynamics of the DP operators which lead to electromagnetic field dissipation in weakly absorptive, structured dielectric media. In doing so we introduce the concept of effective photons as superpositions of the DP excitations which form the natural excitations of the electromagnetic field in this context. As such, the effective photon representation exhibits a clear correspondence with the modes of the non-absorptive regime, while the effects of dissipation manifest in each mode acquiring an effective line-width which leads to decay of the electromagnetic field. Our approach circumvents the need to explicitly specify the initial state of the medium in order to avoid unnatural initial states, and therefore offers a convenient method to include material absorption into treatments of quantum optical processes where it has been neglected.

As an example of its use, at the end of the paper we apply  the formalism to an important case in which dissipation plays a key role---light absorption in structured materials.
The ability of periodic media to enhance numerous optical phenomena \cite{Khurgin:2009} has initiated interest in the form of the Beer-Lambert-Bouguer (BLB) absorption law for photonic structures \cite{Mortensen:2007,Mork:2010}. Here we derive such a result by employing the quantum formalism outlined above and include the effects of material dispersion at all levels of the theory. It is well known that slow light through material dispersion alone cannot serve to enhance the BLB absorption coefficient, while structurally obtained slow light \emph{can} lead to such an enhancement \cite{Mortensen:2007,Chin:2009}. However, in general structured media, the effect of the interplay of structural and material dispersion in this context is not clear. Our expression for the BLB absorption coefficient accounts for both material and structural effects and thus represents a generalisation of the classical treatments presented to date which have neglected material dispersion in the lowest order of approximation.

This paper is structured as follows. In Section \ref{DP} we summarise the quantised theory of macroscopic electrodynamics in dielectric media with particular attention to the DP mode representation. We then obtain approximate results for the DP mode fields in the weakly absorptive dielectric media in Section \ref{WAM} which we then exploit in Section \ref{EP} to introduce the effective photon concept. The dynamics of the effective photon operators are treated in Section \ref{EPD} and applied to the Beer-Lambert-Bouguer law in Section \ref{BL}. We then conclude with a discussion of the results in Section \ref{CR}.

\section{Dressed photons
\label{DP}
}
Constructing the effective photon states requires considerable theoretical apparatus constructed previously by ourselves and many others.  Here we review the results of the complete quantised theory for the macroscopic electromagnetic field in a structured, linear dielectric medium using the DP representation. This section therefore forms a summary of the results presented by Bhat and Sipe \cite{Bhat:2006}, though here we consider an isotropic material for simplicity. However, before reviewing the existing theory we first recall the general principles of Fano theory \cite{Fano:1961} upon which the DP approach to the quantisation of macroscopic electrodynamics is based.
We will thus show that virtually all of the formulae of the dressed photon description have simpler analogs in Fano theory.

Many dissipative quantum systems may be modelled as a primary system of interest coupled to a reservoir in the form of a continuum, where the reservoir often represents the much larger environment of the primary system. Such dissipative quantum systems can be addressed with a number of standard strategies~\cite{Walls:1994,Barnett:1997,Louisell:1973}. In systems where only statistical knowledge of the state is at hand a Lindblad master equation approach is appropriate, while for pure state evolution a Weisskopf-Wigner treatment may be applied. In carrying this latter approach over to the Heisenberg picture, the Markov approximation may be employed, where appropriate, in dealing with the effect of the continuum on the bare states of the primary system to obtain evolution equations of the Langevin type. Finally, we have Fano theory \cite{Barnett:1997} in which the diagonalisation of the full Hamiltonian describing both primary system and reservoir is pursued. A key attraction of this approach is that the diagonalising states thus obtained evolve harmonically in time and provide a convenient basis for describing the dynamics of the system.

\subsection{Fano theory: simple model\label{fanosimple}
}
In its simplest form Fano theory consists of determining the eigenmodes, or dressed states, of a system of harmonic oscillators in which a single discrete state is coupled to a continuum. Such a treatment may be found in standard texts on quantum optics (e.g. \cite{Barnett:1997}) in relation to dissipative systems (e.g. a single cavity mode coupled to the radiation modes of free space). Here we present the Fano treatment of this simple system to provide a familiar reference with which to compare the results presented later for macroscopic electrodynamics.

In the Heisenberg picture we therefore begin with a Hamiltonian of the form
\begin{equation}
\Hamop=\Hamop_{\rm dis}+\Hamop_{\rm con}+\Hamop_{\rm int},
\label{Hexform}
\end{equation}
where $\Hamop_{\rm dis}$, $\Hamop_{\rm con}$, and $\Hamop_{\rm int}$ are the discrete state, continuum, and bilinear interaction Hamiltonians, respectively. Such a Hamiltonian may be written explicitly as
\begin{eqnarray}
\Hamop&=&\hbar\om_1\aop^\dagger\aop+\int^\infty_0\dom\hbar\om\bop^\dagger_\om\bop_\om\nonumber\\
&&+\hbar\sqrt{\frac{\om_1}{2}}\int^\infty_0\dom \Lambda(\om)\left[\aop+\aop^\dagger\right]\left[\bop_\om+\bop^{\dagger}_\om\right],
\label{Hex}
\end{eqnarray}
where the prefactors preceding the integral in the interaction term represent a convenient scaling, $^\dagger$ denotes a Hermitian adjoint, $\Lambda(\om)$ is a real, dimensionless coupling strength, and $\om_1$ and $\om$ are the frequencies of the bare discrete and continuum states, respectively. The creation and annihilation operators for these states obey the equal-time commutation relations (ETCRs)
\begin{eqnarray}
\left[\aop,\aop^\dagger\right]=1,\label{aacr1}\\
\left[\bop_\om,\bop_{\om'}^\dagger\right]=\delta(\om-\om'),\label{bbcr1}
\end{eqnarray}
where $\delta(\om)$ is the Dirac delta distribution. The Fano diagonalisation of the Hamiltonian (\ref{Hex}) results in
\begin{equation}
\Hamop=\int_0^\infty\dom\hbar\om\cop^\dagger_\om\cop_\om,
\label{Hexdiag}
\end{equation}
where the continuum of dressed operators satisfy the eigenvalue equation
\begin{equation}
\hbar\om\cop_\om=\left[\cop_\om,\Hamop\right],
\label{ceq}
\end{equation}
and are given in terms of the bare operators as
\begin{equation}
\cop_\om=\alpha(\om)\aop+\bar{\alpha}(\om)\aopa+\int^\infty_0\dom'\left[\beta(\om,\om')\bop_{\om'}+\bar{\beta}(\om,\om')\bopa_{\om'}\right].
\label{cex}
\end{equation}
The coefficients $\alpha(\om)$, $\bar{\alpha}(\om)$, $\beta(\om,\om')$ and $\bar{\beta}(\om,\om')$ are determined by inserting (\ref{cex}) into (\ref{ceq}) and using the ETCRs to obtain
\begin{eqnarray}
\left(\om-\om_1\right)\alpha(\om)=\sqrt{\frac{\om_1}{2}}\int^\infty_0\dom'\Lambda(\om')\left[\beta(\om,\om')-\bar{\beta}(\om,\om')\right],
\label{alphaeq}\\
\left(\om+\om_1\right)\bar{\alpha}(\om)=\sqrt{\frac{\om_1}{2}}\int^\infty_0\dom'\Lambda(\om')\left[\beta(\om,\om')-\bar{\beta}(\om,\om')\right],
\label{alphabeq}\\
\left(\om-\om'\right)\beta(\om,\om')=\sqrt{\frac{\om_1}{2}}\Lambda(\om')\left[\alpha(\om)-\bar{\alpha}(\om)\right],
\label{betaeq}\\
\left(\om+\om'\right)\bar{\beta}(\om,\om')=\sqrt{\frac{\om_1}{2}}\Lambda(\om')\left[\alpha(\om)-\bar{\alpha}(\om)\right].
\label{betabeq}
\end{eqnarray}
Inspection of (\ref{alphaeq}) and (\ref{alphabeq}) yields $(\om+\om_1)\bar{\alpha}(\om)=(\om-\om_1)\alpha(\om)$, while solving (\ref{betaeq}) and (\ref{betabeq}) for $\beta(\om,\om')$ and $\bar{\beta}(\om,\om')$ leads to
\begin{eqnarray}
\beta(\om,\om')=\left[P\frac{1}{\om-\om'}+z_\om\delta(\om-\om')\right]\Lambda(\om')\gamma(\om),\label{betasol}\\
\bar{\beta}(\om,\om')=\frac{1}{\om+\om'}\Lambda(\om')\gamma(\om),\label{betabsol}
\end{eqnarray}
where $P$ indicates that the Cauchy principal value is to be taken upon integration and we have introduced $\gamma(\om)=\alpha(\om)-\bar{\alpha}(\om)$. The quantity $z_\om$ originates from the treatment of the singularity at $\om=\om'$ in (\ref{betasol}) using a technique introduced by Dirac \cite{Dirac:1927} and is a hallmark of the Fano approach. It parametrises the resonant contribution of the continuum operators to the dressed operators and is determined by the consistency of (\ref{alphaeq})--(\ref{betabeq}). We now multiply (\ref{alphaeq}) and (\ref{alphabeq}) by $\om+\om_1$ and $\om-\om_1$, respectively, add the resulting expressions and use (\ref{betasol}) and (\ref{betabsol}) to obtain
\begin{eqnarray}
&&\om_1^2\left\{1-P\int^\infty_0\dom'\frac{\left[\Lambda(\om')\right]^2}{\om'-\om}-\int^\infty_0\dom'\frac{\left[\Lambda(\om')\right]^2}{\om'+\om}+z_\om\left[\Lambda(\om)\right]^2\right\}\gamma(\om)\nonumber\\
&=&\om^2\gamma(\om).
\label{gammaeq1}
\end{eqnarray}
Defining the function $\Gamma(\om)$ through its real and imaginary parts as
\begin{eqnarray}
\real\Gamma(\om)=P\int^\infty_{-\infty}\dom'\frac{\om'}{|\om'|}\frac{\left[\Lambda(|\om'|)\right]^2}{\om'-\om},\label{rGammadef}\\
\imag\Gamma(\om)=\pi\frac{\om}{|\om|}\left[\Lambda(|\om|)\right]^2,\label{iGammadef}
\end{eqnarray}
we may rewrite (\ref{gammaeq1}) in the form
\begin{equation}
\om_1^2\left[1-\real\Gamma(\om)+\frac{z_\om}{\pi}\imag\Gamma(\om)\right]\gamma(\om)=\om^2\gamma(\om).
\label{gammaeq2}
\end{equation}
The results (\ref{rGammadef})--(\ref{gammaeq2}) demonstrate that the relationship between $z_\om$ and the corresponding dressed state frequency $\om$ is mediated by the causal function $\Gamma(\om)$ which satisfies the standard Kramers-Kronig relations \cite{Landau:1960}
\begin{eqnarray}
\real\Gamma(\om)=\frac{1}{\pi}P\int^\infty_{-\infty}\dom'\frac{\imag\Gamma(\om')}{\om'-\om},
\label{KKri}\\
\imag\Gamma(\om)=-\frac{1}{\pi}P\int^\infty_{-\infty}\dom'\frac{\real\Gamma(\om')}{\om'-\om}.
\label{KKir}
\end{eqnarray}
The result for $z_\om$ then follows as
\begin{equation}
z_\om=\pi\frac{\om^2-\om_1^2\left[1-\real\Gamma(\om)\right]}{\om_1^2\imag\Gamma(\om)},
\label{zsol}
\end{equation}
which implies that $z_\om$ is real. Finally, the solution for $\gamma(\om)$ is determined by imposing the ETCR
\begin{equation}
\left[\cop_\om,\copa_{\om'}\right]=\delta(\om-\om').
\label{ccetcr}
\end{equation}
Inserting (\ref{cex}) into (\ref{ccetcr}) and using (\ref{gammaeq2}) then yields
\begin{equation}
\frac{\om_1}{2}\imag\Gamma(\om)|\gamma(\om)|^2=\frac{\pi}{z^2_\om+\pi^2},
\label{gamma2sol}
\end{equation}
from which we obtain
\begin{equation}
\gamma(\om)=\sqrt{\frac{2}{\om_1}}\frac{1}{\Lambda(\om)\left(z_\om+\rmi\pi\right)},
\label{gammasol}
\end{equation}
where (\ref{iGammadef}) and the freedom to choose the arbitrary phase implicit in (\ref{gamma2sol}) have been exploited. Using the expansion (\ref{cex}) and the ETCRs we may now construct the bare discrete state operator as
\begin{equation}
\aop(t)=\int^\infty_0 \dom \left[\alpha^*(\om)\cop_\om(t)-\bar{\alpha}(\om)\copa_\om(t)\right],
\label{aex}
\end{equation}
where
\begin{eqnarray}
\alpha(\om)=\frac{\om+\om_1}{2\om_1}\gamma(\om),
\label{alphasol}\\
\bar{\alpha}(\om)=\frac{\om-\om_1}{2\om_1}\gamma(\om).
\label{alphabsol}
\end{eqnarray}
Thus, combining (\ref{zsol}) and (\ref{gammasol})--(\ref{alphabsol}) along with $\cop_\om(t)=\cop_\om(0)\exp(-\rmi\om t)$, the dynamics of $\aop$ may be established. For a system where only the discrete state is excited initially, the dissipation of energy in this mode manifests through the dephasing of the harmonically varying dressed states in (\ref{aex}).

To summarise, a choice of the bare discrete state frequency $\omega_1$ and the response function $\Gamma(\omega)$ (which sets the coupling $\Lambda(\omega)$), determines how the discrete state contributes to each member of the new dressed state continuum.  For each dressed state frequency $\omega$, the eigenvalue $z_\omega$ quantifies this contribution. 

\subsection{Fano theory: dressed photons in macroscopic electrodynamics\label{fanoem}
}
We now commence a review of the results presented by Bhat and Sipe \cite{Bhat:2006} which extends the Fano description to the problem of causal dissipative macroscopic electrodynamics. The Hamiltonian governing the linear interaction between the vacuum electromagnetic (photon) field and an inhomogeneous medium may be written, in analogy with (\ref{Hexform}), in the form
\begin{equation}
\Hamop=\Hamop_{\rm phot}+\Hamop_{\rm med}+\Hamop_{\rm int},
\label{Hform}
\end{equation}
where $\Hamop_{\rm phot}$, $\Hamop_{\rm med}$ and $\Hamop_{\rm int}$ are the photon field, medium, and bilinear interaction Hamiltonians, respectively. However, due to the addition of spatial degrees of freedom, we now have the usual set of vacuum field oscillators coupled to a continuum of medium oscillators for each spatial dimension at each field point $\rv$. Although the situation is therefore more complicated, essentially the same principles apply here as in Section \ref{fanosimple}. The Hamiltonian (\ref{Hform}) may be written explicitly as
\begin{eqnarray}
\Hamop&=&\frac{1}{2}\int\drv\left[\frac{\Dop(\rv)\cdot\Dop(\rv)}{\varepsilon_0}+\frac{\Bop(\rv)\cdot\Bop(\rv)}{\mu_0}\right]\nonumber\\
&&+\int\drv\int_0^\infty\dom\hbar\om\bopva_\om(\rv)\cdot\bopv_\om(\rv)\nonumber\\
&&-\sqrt{\frac{\hbar}{\varepsilon_0}}\int_0^\infty\dom\int\drv\Lambda(\rv,\om)\Dop(\rv)\cdot\left[\bopv_\om(\rv)+\bopva_\om(\rv)\right],
\label{Hdef}
\end{eqnarray}
where $\varepsilon_0$ is the permittivity and $\mu_0$ the permeability of free space, $\Lambda(\rv,\om)$ is again a real, dimensionless coupling strength, and the various operators satisfy the ETCRs
\begin{eqnarray}
\left[\hat{D}_i(\rv),\hat{B}_j(\rv')\right]=i\hbar\epsilon_{ilj}\frac{\partial}{\partial r_l}\delta(\rv-\rv'),
\label{DBcr}\\
\left[\hat{b}_{i\om}(\rv),\hat{b}^\dagger_{j\om'}(\rv')\right]=\delta_{ij}\delta(\om-\om')\delta(\rv-\rv').
\label{bbcr}
\end{eqnarray}
Here Latin indices denote Cartesian vector components, repeated indices imply a summation, $\delta_{ij}$ is the Kronecker delta, and $\epsilon_{ilj}$ are the components of the Levi-Civita pseudotensor. A Fano diagonalisation of $\Hamop$ yields
\begin{equation}
\Hamop=\sum_m\inth\dom\hbar\om\copa_{m\om}\cop_{m\om},
\label{Hdiag}
\end{equation}
where $\copa_{m\om}$ and $\cop_{m\om}$ are, respectively, the creation and annihilation operators for the collective DP modes with frequency $\om$. 
The discrete mode index $m$ arises due to the additional degrees of freedom in the system (\ref{Hdef}), in contrast to the case (\ref{Hexdiag}) where only a single discrete state is involved. We note, however, that this discrete nature of the mode index $m$ is simply a notational convenience and it may in practice represent a combination of discrete and continuous indices as the geometry of the medium dictates (e.g. in the case of a one dimensional photonic crystal structure, as considered in Section \ref{BL}, we have $m\to(\sigma,k)$, where $\sigma$ is the discrete band index and $k$ the continuous wavevector in the direction of periodicity). The DP operators satisfy the ETCRs
\begin{equation}
\left[\cop_{m\om},\copa_{m'\om'}\right]=\delta_{mm'}\delta(\om-\om'),
\label{cccr}
\end{equation}
and evolve in time according to
\begin{equation}
\cop_{m\om}(t)=\cop_{m\om}(0)\exp\left(-\rmi\om t\right).
\label{copt}
\end{equation}

Note that only those collective excitations which involve contributions from the transverse photon modes have been retained in (\ref{Hdiag}) since these are the only objects of interest here: the remaining longitudinal modes do not involve couplings to the transverse electromagnetic field and thus do not participate in relevant optical processes.

In the dipole approximation implicit to the theory presented here, the electric displacement operator $\Dop(\rv)$ represents the momentum conjugate to the vector potential operator $\hat{\bi{A}}(\rv)$ \cite{Power:1959} and is therefore a convenient field variable to work with. This is especially so in extensions to nonlinear media \cite{Yang:2008}. In terms of the DP operators the electric displacement operator is constructed in analogy to (\ref{aex}) as
\begin{equation}
\Dop(\rv)=\sum_m\inth\dom\Dv_{m\om}(\rv)\cop_{m\om}+H.a.,
\label{Dopex}
\end{equation}
where $H.a$ represents the Hermitian adjoint of the preceding terms and the coefficients $\Dv_{m\om}(\rv)$ are the DP mode fields. As described for the single discrete state in Section \ref{fanosimple}, dissipation of the electromagnetic field arises due to dephasing of some initial superposition of the harmonically varying DP modes. These are obtained as solutions to equation (118) in \cite{Bhat:2006} which is a direct analogue to (\ref{gammaeq2}) in the Fano diagonalisation process, viz.,
\begin{equation}
\nabv\times\nabv\times\left\{\left[1-\real\Gamma(\rv,\om)+\frac{z_{m\om}}{\pi}\imag\Gamma(\rv,\om)\right]\Dv_{m\om}(\rv)\right\}=\frac{\om^2}{c^2}\Dv_{m\om}(\rv),
\label{Dmeq}
\end{equation}
where $c$ is the vacuum speed of light, and $\real\Gamma(\rv,\om)+\rmi\imag\Gamma(\rv,\om)=\Gamma(\rv,\om)=1-1/\epsilon(\rv,\om)$, with $\epsilon(\rv,\om)$ the relative linear permittivity from the classical theory, which actually defines the problem. That is, given $\epsilon(\rv,\om)$ (or $\Gamma(\rv,\om)$) as the sole input, the DP mode fields may then be obtained from (\ref{Dmeq}) and used to construct $\Dop$ through (\ref{Dopex}). As in Section \ref{fanosimple}, the real and imaginary parts of $\Gamma(\rv,\om)$ are related to the coupling strength $\Lambda(\rv,\om)$ through (cf. (\ref{rGammadef}) and (\ref{iGammadef}))
\begin{eqnarray}
\real\Gamma(\rv,\om)=P\int^\infty_{-\infty}\dom'\frac{\om'}{|\om'|}\frac{\left[\Lambda(\rv,|\om'|)\right]^2}{\om'-\om},\label{rGammadef1}\\
\imag\Gamma(\rv,\om)=\pi\frac{\om}{|\om|}\left[\Lambda(\rv,|\om|)\right]^2.\label{iGammadef1}
\end{eqnarray}
and therefore satisfy the Kramers-Kronig relations (cf. (\ref{KKri}) and (\ref{KKir}))
\begin{eqnarray}
\real\Gamma(\rv,\om)=\frac{1}{\pi}P\intf\dom'\frac{\imag\Gamma(\rv,\om')}{\om'-\om},
\label{KKri1}\\
\imag\Gamma(\rv,\om)=-\frac{1}{\pi}P\intf\dom'\frac{\real\Gamma(\rv,\om')}{\om'-\om},
\label{KKir1}
\end{eqnarray}
where $\imag\Gamma(\rv,\om)>0$ for $\om>0$. Once the DP mode fields have been obtained using (\ref{Dmeq}), the commutation relation (\ref{cccr}) may be imposed in analogy with the derivation of (\ref{gamma2sol}) to obtain the normalisation condition
\begin{equation}
\frac{\pi}{\epsilon_0\hbar}\int\drv\;\imag\Gamma(\rv,\om)\Dv^*_{m\om}(\rv)\cdot\Dv_{m\om}(\rv)=\frac{\pi^2}{z_{m\om}^2+\pi^2}.
\label{Dmnorm}
\end{equation}

Due to the increased complexity of the present system with respect to that presented in Section \ref{fanosimple}, the determination of $z_{m\om}$ is somewhat more involved: it has been shown \cite{Bhat:2006} that the mode equation (\ref{Dmeq}) may be recast as a generalised Hermitian system involving $\Gamma(\rv,\om)$ and $\om$ as inputs, for which $z_{m\om}$ is the real eigenvalue. Thus, for each frequency $\om$ the set of mode solutions indexed by $m$ may be shown to constitute a complete set. The full set of DP mode fields across all frequencies is therefore highly over-complete. This property reflects the multitude of possible states associated with the medium degrees of freedom which, at any instant in time, can manifest the same spatial electromagnetic field distribution. We note that for solutions with $z_{m\om}=0$ the result (\ref{Dmeq}) recovers the familiar non-Hermitian mode equation from standard linear optics in structured, lossless dispersive media. Indeed, we shall find in Section \ref{WAM} that such solutions are of central importance to the description of the electromagnetic field in weakly absorptive media.

One can also find expressions corresponding to (\ref{Dopex}) for the other electromagnetic field variables. 
As they are not needed for the remaining derivations, we reserve these for the Appendix.
In summary, the results (\ref{Hdiag})--(\ref{Dmeq}) and (\ref{Dmnorm}), together with (\ref{Eopex}), (\ref{Bopex}) and (\ref{Bmdef}) from the Appendix constitute a complete quantised description of the macroscopic electromagnetic field in a causal dielectric medium.

\section{Weakly absorptive media
\label{WAM}}
With the Fano theory established, we now turn to the effective temporal evolution of electromagnetic states in a dissipative medium.  So as to  establish a correspondence with the non-absorptive regime we now search for approximate solutions for the DP mode properties in the limit of weak absorption. We consider a medium that is weakly absorptive in the sense that, over a frequency interval $\Delta\om$ centred at $\bar\om$, we may assume that $\imag\Gamma(\rv,\om)\ll\Delta\om/\bar\om\ll 1$; this ensures that any anticipated line-shape function of width $\sim\bar\om\imag\Gamma(\rv,\bar\om)$ will be well contained on the interval $\Delta\om$. Furthermore, in order to pursue a perturbation approach with regard to the mode equation (\ref{Dmeq}) we assume that for a particular mode index $m$ we may consider the absorptive part of the medium response as satisfying $|z_{m\om}\imag\Gamma(\rv,\om)|/\pi\sim\Delta\om/\bar\om$, and that the frequency variation of $\Gamma(\rv,\om)$ over the frequency interval $\Delta\om$ may likewise be considered a perturbative effect of the same order. We then introduce the formal expansions
\begin{eqnarray}
z_{m\om}=z^{(0)}_{m\om}+z^{(1)}_{m\om}+z^{(2)}_{m\om}+\ldots,
\label{zexpert}\\
\Dv_{m\om}(\rv)=\Dv_{m\om}^{(0)}(\rv)+\Dv_{m\om}^{(1)}(\rv)+\Dv_{m\om}^{(2)}(\rv)+\ldots,
\label{Dexpert}
\end{eqnarray}
where the superscripts denote the order of each term in the expansion with respect to the perturbative terms. Collecting lowest order terms in (\ref{Dmeq}) we obtain
\begin{equation}
\nabv\times\nabv\times\left\{\left[1-\real\Gamma(\rv,\bar\om)\right]\Dv_{m\om}^{(0)}(\rv)\right\}=\frac{\bar\om^2}{c^2}\Dv^{(0)}_{m\om}(\rv).
\label{Deq0}
\end{equation}
Comparison with (\ref{Dmeq}) implies that $z_{m\bar\om}=0$ and that the spatial form of $\Dv^{(0)}_{m\om}(\rv)$ is independent of $\om$. However, in order to be consistent with the normalisation condition (\ref{Dmnorm}) to lowest order we require
\begin{equation}
\frac{\pi}{\epsilon_0\hbar}\int\drv\;\imag\Gamma(\rv,\bar\om)\left[\Dv^{(0)}_{m\om}(\rv)\right]^*\cdot\Dv^{(0)}_{m\om}(\rv)=\frac{\pi^2}{\left[z^{(0)}_{m\om}\right]^2+\pi^2}.
\label{Dmnorm0}
\end{equation}

Now consider the generalised Hermitian eigenvalue problem
\begin{eqnarray}
&&\left[1-\real\Gamma(\rv,\bar\om)\right]\nabv\times\nabv\times\left\{\left[1-\real\Gamma(\rv,\bar\om)\right]\tilde{\Dv}_{\tilde{\om}}(\rv)\right\}\nonumber\\
&=&\frac{\tilde{\om}^2}{c^2}\left[1-\real\Gamma(\rv,\bar\om)\right]\tilde{\Dv}_{\tilde{\om}}(\rv),
\label{Deq0h}
\end{eqnarray}
where the argument $\bar\om$ of the medium response is fixed independently of the eigenvalue $\tilde{\om}^2/c^2$. This corresponds to a structured medium in the absence of material dispersion and loss, and recalling (\ref{Deq0}) we observe that $\Dv^{(0)}_{m\om}(\rv)$ is a solution to (\ref{Deq0h}) with $\tilde{\om}=\bar\om$. As such, the corrections to this lowest order solution may be calculated using standard perturbation theory. Collecting first order terms in (\ref{Dmeq}) as
\begin{eqnarray}
&&\nabv\times\nabv\times\left\{\left[1-\real\Gamma(\rv,\bar\om)\right]\Dv_{m\om}^{(1)}(\rv)\right\}\nonumber\\
&&+\nabv\times\nabv\times\left\{\left[\frac{z^{(0)}_{m\om}}{\pi}\imag\Gamma(\rv,\bar\om)-\left(\om-\om_m\right)\left.\frac{\partial\real\Gamma(\rv,\om)}{\partial\om}\right\vert_{\bar\om}\right]\Dv^{(0)}_{m\om}(\rv)\right\}\nonumber\\
&=&\frac{\bar\om^2}{c^2}\Dv^{(1)}_{m\om}(\rv)+\frac{2\bar\om(\om-\bar\om)}{c^2}\Dv^{(0)}_{m\om}(\rv).
\label{Deq1}
\end{eqnarray}
we may multiply this expression by $\left[1-\real\Gamma(\rv,\bar\om)\right]$ and exploit the Hermitian nature of (\ref{Deq0h}) to eliminate the terms involving $\Dv^{(1)}_{m\om}(\rv)$, leaving us with
\begin{eqnarray}
&&\int\drv\left[\frac{z^{(0)}_{m\om}}{\pi}\imag\Gamma(\rv,\bar\om)-\left(\om-\bar\om\right)\left.\frac{\partial\real\Gamma(\rv,\om)}{\partial\om}\right\vert_{\bar\om}\right]\nonumber\\
&&\times\left[\Dv^{(0)}_{m\om}(\rv)\right]^*\cdot\Dv^{(0)}_{m\om}(\rv)\nonumber\\
&=&\frac{2\bar\om(\om-\bar\om)}{c^2}\int\drv\left[1-\real\Gamma(\rv,\bar\om)\right]\left[\Dv^{(0)}_{m\om}(\rv)\right]^*\cdot\Dv^{(0)}_{m\om}(\rv),
\label{Deq11}
\end{eqnarray}
where (\ref{Deq0}) has been used and we have assumed that the mode fields either vanish or satisfy periodic conditions at the spatial boundaries in order to remove the boundary terms that appear in the rearrangement of the integral over $\rv$ on the left hand side of (\ref{Deq11}). We note that, although degeneracy in the solutions to (\ref{Deq0h}) is to be expected in general, we treat the non-degenerate case here for simplicity. Where symmetry considerations fail to expedite the issue, the calculation for the degenerate case may be performed straightforwardly in the usual way \cite{Merzbacher:1970}. Rearranging (\ref{Deq11}) then yields an expression for $z^{(0)}_{m\om}$ of the form
\begin{equation}
z^{(0)}_{m\om}=\pi\frac{\om-\bar\om}{\Delta_m}.
\label{z0sol}
\end{equation}
where
\begin{equation}
\Delta_m=\frac{\bar\om\int\drv\imag\Gamma(\rv,\bar\om)\left[\Dv^{(0)}_{m\om}(\rv)\right]^*\cdot\Dv^{(0)}_{m\om}(\rv)}{2\int\drv\left[1-\real\Gamma(\rv,\bar\om)+\frac{\bar\om}{2}\left.\frac{\partial\real\Gamma(\rv,\om)}{\partial\om}\right\vert_{\bar\om}\right]\left[\Dv^{(0)}_{m\om}(\rv)\right]^*\cdot\Dv^{(0)}_{m\om}(\rv)}.
\label{Deltadef}
\end{equation}
In considering weakly dissipative media we may expect the inequality $1-\real\Gamma(\rv,\om)+\om[\partial\real\Gamma(\rv,\om)/\partial\om]/2>0$ to hold over the frequency range of interest \cite{Landau:1960} which implies that $\Delta_m>0$.

We shall find it convenient to allow the reference frequency to vary with the mode index $m$ by setting $\bar\om=\om_m$, and to define $\Dv_m(\rv)$ as a solution of
\begin{equation}
\nabv\times\nabv\times\left\{\left[1-\real\Gamma(\rv,\om_m)\right]\Dv_{m}(\rv)\right\}=\frac{\om_m^2}{c^2}\Dv_{m}(\rv).
\label{Deqm}
\end{equation}
From~(\ref{Deq0}), we can assert $\Dv^{(0)}_m(\rv)=C\Dv_m(\rv)$ for some constant
$C$, so that we may rewrite (\ref{Deltadef}) in terms of $\Dv_m(\rv)$ and $\om_m$ as
\begin{equation}
\Delta_m=\frac{\om_m\int\drv\imag\Gamma(\rv,\om_m)\Dv^*_{m}(\rv)\cdot\Dv_{m}(\rv)}{2\int\drv\left[1-\real\Gamma(\rv,\om_m)+\frac{\om_m}{2}\left.\frac{\partial\real\Gamma(\rv,\om)}{\partial\om}\right\vert_{\om_m}\right]\Dv^*_{m}(\rv)\cdot\Dv_{m}(\rv)}.
\label{Deltadef1}
\end{equation}
Furthermore, normalising $\Dv_m(\rv)$ according to
\begin{eqnarray}
&&\frac{1}{\epsilon_0}\int\drv\left[1-\real\Gamma(\rv,\om_m)+\frac{\om_m}{2}\left.\frac{\partial\real\Gamma(\rv,\om)}{\partial \om}\right\vert_{\om_m}\right]\Dv^*_m(\rv)\cdot\Dv_m(\rv)\nonumber\\
&=&\frac{\hbar\om_m}{2}.
\label{Dmenorm}
\end{eqnarray}
allows us to write the lowest order mode field solutions as
\begin{equation}
\Dv^{(0)}_{m\om}(\rv)=\sqrt{\frac{\pi}{\Delta_m}}\frac{1}{z^{(0)}_{m\om}+\rmi\pi}\Dv_{m}(\rv),
\label{D0norm}
\end{equation}
Note that in writing (\ref{D0norm}) we have fixed the arbitrary phase implicit in (\ref{Dmnorm0}) as was done in obtaining (\ref{gammasol}) from (\ref{gamma2sol}). Concluding the perturbation treatment here, we note our key results to be the lowest order solution for $\Dv_{m\om}(\rv)$ given by (\ref{D0norm}), and that for $z_{m\om}$ given by (\ref{z0sol}) and (\ref{Deltadef1}).

\section{Effective photons
\label{EP}
}
We now exploit the results of Section \ref{WAM} to derive useful expressions for the electromagnetic field operators in the weak absorption regime. Inserting (\ref{D0norm}) into (\ref{Dopex}) with (\ref{z0sol}) and (\ref{Deltadef}) we obtain an approximate expression for the electric displacement operator as
\begin{equation}
\Dop(\rv,t)\simeq\sum_m\Dv_{m}(\rv)\sqrt{\frac{\Delta_m}{\pi}}\inth\dom\frac{1}{\om-\om_m+\rmi\Delta_m}\cop_{m\om}(t)+H.a.,
\label{Dopapprox}
\end{equation}
where we again emphasise that the sum over $m$ may imply the combination of a sum and integral over discrete and continuous band indices, respectively. Recalling the exact expression (\ref{Dopex}), the salient observation to be made here is the separation of the spatial variation of the mode fields from the frequency integral in (\ref{Dopapprox}). To this order of approximation, the departure of the DP modes from the non-absorptive solutions affects only the magnitude and phase of the mode fields while leaving their spatial form unchanged (see (\ref{D0norm})). It is this crucial property of the weak absorption limit that, through (\ref{Dopapprox}), ascribes a natural superposition of DP excitations to a single field distribution, and thus reduces enormously the number of mode fields required to describe the electromagnetic field. This motivates the definition of a new class of bosons in the form of composite DPs which we term effective photons (EPs). The operators corresponding to these excitations are defined as
\begin{equation}
\aop_m(t)=\sqrt{\frac{\Delta_m}{\pi}}\inth\dom\frac{1}{\om-\om_m+\rmi\Delta_m}\cop_{m\om}(t),
\label{aopdef}
\end{equation}
where the scaling factor has been chosen for convenience, and we observe that $\Delta_m$ enters here as the EP line-width governing the magnitude and phase of the contributions from the various DP operators. It is this uncertainty in the EP frequency which leads to dissipation in the same spirit as in the spontaneous emission of an atom, though here the roles of matter and radiation are reversed. The EP operators, through (\ref{cccr}), satisfy the ETCR
\begin{equation}
\left[\aop_m,\aopa_{m'}\right]=\delta_{mm'}\frac{\Delta_m}{\pi}\inth\dom\frac{1}{(\om-\om_m)^2+\Delta^2_m}.
\label{aacr}
\end{equation}
We recall that by assumption the range of validity of the perturbation treatment in Section \ref{WAM} ensures $\Delta\om\gg\om_m\imag\Gamma(\rv,\om_m)$ and, through (\ref{Deltadef}), we then have $\Delta_m\sim\om_m\imag\Gamma(\rv,\om_m)$ and thus $\Delta_m\ll\Delta\om$. This frequency range therefore includes the dominant contribution to the integral on the right hand side of (\ref{aacr}) arising from a narrow region about $\om=\om_m$. Extending the domain of integration to encompass the entire real frequency axis we may then obtain the approximate relation
\begin{equation}
\left[\aop_m,\aopa_{m'}\right]\simeq\delta_{mm'}.
\label{aacrapprox}
\end{equation}
Substituting (\ref{aopdef}) into (\ref{Dopapprox}) we then obtain the following expression for the electric displacement operator,
\begin{equation}
\Dop(\rv)\simeq\sum_m\Dv_m(\rv)\aop_m+H.a..
\label{Dopa}
\end{equation}
To the same lowest order we also have from (\ref{Eopex}), (\ref{Bopex}) and (\ref{Bmdef}) in the Appendix
\begin{eqnarray}
\Eop(\rv)\simeq\frac{1}{\epsilon_0}\sum_m\left[1-\real\Gamma(\rv,\om_m)\right]\Dv_m(\rv)\aop_m+H.a.,\label{Eopa}\\
\Bop(\rv)\simeq-\sum_m\frac{\rmi}{\epsilon_0\om_m}\nabv\times\left\{\left[1-\real\Gamma(\rv,\om_m)\right]\Dv_m(\rv)\right\}\aop_m+H.a.,
\label{Bopa}
\end{eqnarray}
with, again, $\Hop(\rv)=\Bop(\rv)/\mu_0$.

The approximate expressions (\ref{aacrapprox})--(\ref{Bopa}) in the weak absorption regime are observed to be identical in form to the corresponding exact results for photons in vacuum \cite{Berestetski:1984} and DPs in perfectly transparent media \cite{Bhat:2006,Sipe:2009}. Of course, in the present context the EP operators do not diagonalise the Hamiltonian which is given instead by (\ref{Hdiag}). A consequence of this is the non-vanishing EP line-width which now follows from (\ref{Deltadef1}) and (\ref{Dmenorm}) as
\begin{equation}
\Delta_m=\frac{1}{\epsilon_0\hbar}\int\drv\imag\Gamma(\rv,\om_m)\Dv_m^*(\rv)\cdot\Dv_m(\rv).
\label{Deltadefnorm}
\end{equation}
It is this property of the EP operators that provides the key point of difference from the non-absorptive case.

\section{Effective photon dynamics
\label{EPD}
}
For the purpose of determining the dynamics of the EP operators it is useful to establish  from the definition (\ref{aopdef}) the unequal time commutation relations (UTCRs)
\begin{equation}
\left[\aop_m(t),\aopa_{m'}(t')\right]\simeq\delta_{mm'}\exp\left[-\rmi\om_m(t-t')-\Delta_m|t-t'|\right],
\label{aacrn}
\end{equation}
and
\begin{equation}
\left[\cop_{m\om}(t),\aopa_{m'}(t')\right]=\delta_{mm'}\sqrt{\frac{\Delta_m}{\pi}}\frac{\exp\left[-\rmi\om(t-t')\right]}{\om-\om_m-\rmi\Delta_m}.
\label{cacrn}
\end{equation}
The result (\ref{aacrn}), from which the dynamics of all electromagnetic field quantities corresponding to EP states may be established, is an approximate relation obtained via the same procedure leading to (\ref{aacrapprox}).

To demonstrate the effects of dissipation upon the dynamics of the EP operators we consider the EP coherent states of the form \cite{Yang:2008}
\begin{equation}
|\alpha\rangle=\exp\left(-\frac{|\alpha^2|}{2}\right)\sum^\infty_{n=0}\frac{\alpha^n}{n!}\left\{\sum_m\phi_m\aopa_m(0)\right\}^n|vac\rangle,
\label{psidef}
\end{equation}
where $\alpha$ and $\phi_m$ are complex numbers and $\sum_m|\phi_m|^2=1$. The expectation value of the EP annihilation operator $\aop_m$ then follows from (\ref{aacrn}) and (\ref{psidef}) as
\begin{equation}
\langle\aop_m\rangle\simeq\alpha\phi_m\exp\left[-\rmi\om_mt-\Delta_m|t|\right],
\label{aevcs}
\end{equation}
where $\langle\aop_m\rangle=\langle\alpha|\aop_m|\alpha\rangle$. We note that (\ref{aevcs}) is only valid for $|t|\gtrsim1/\Delta\om$ due to the inaccuracy of the approximations applied far from $\om_m$ in the frequency domain. In the region of validity the right hand side of (\ref{aevcs}) decays exponentially away from the time origin, with $\tau_m=1/(2\Delta_m)$ playing the role of an EP lifetime for the mode $m$. Since we are considering a closed Hamiltonian system, the evolution of $\langle\aop_m\rangle$ is bounded at all times and vanishes exponentially as $t\to\pm\infty$. In a more realistic situation, asymmetric time evolution must be included artificially through a transient interaction with an external system.

We have shown that by expressing the electromagnetic field in terms of EP states we circumvent the need to specify the state of the medium explicitly in order to determine the dynamics of the electromagnetic field expectation values. A key advantage here is that the mode fields associated with the EP excitations account for the structural and material dispersion of the medium, and therefore form the natural language for describing the electromagnetic field in contexts where those properties are of importance, such as the study of dispersive photonic crystals and metamaterials.

\section{The Beer-Lambert-Bouguer law in structured media
\label{BL}
}
An application of the theory presented here is the calculation of the decay rate associated with the BLB law which describes the exponential spatial decay of the electromagnetic field intensity within a medium under constant illumination. We note that although this result could in principle be derived classically, that does not so far appear to have happened in full generality. We thus consider it an interesting example of how the more general quantum formalism can help to identify classical results. 

\subsection{The absorption coefficient}
In order to examine propagation in space for this purpose, we consider a structured medium which is periodic in the $x$-direction with period $L_{cell}$ (e.g. a defect waveguide within a 2-D photonic crystal). In this case the mode index $m$ takes the form $(\sigma,k)$, where $\sigma$ is a discrete band index while the Bloch mode wavevector $k$ is continuous. The coherent state expectation value for the displacement field $\langle\Dop(\rv)\rangle=\langle\alpha|\Dop(\rv)|\alpha\rangle$ is then written as
\begin{equation}
\langle\Dop(\rv)\rangle\simeq\sum_\sigma\intf\dk\dv_{\sigma k}(\rv)\langle\aop_{\sigma k}\rangle\exp\left(\rmi kx\right)+c.c.,
\label{Dex}
\end{equation}
where we have introduced the Bloch modes $\dv_{\sigma k}(\rv)$ through $\Dv_{\sigma k}(\rv)=\dv_{\sigma k}(\rv)\exp\left(\rmi kx\right)$. The normalisation which previously involved an integral over all space (cf. (\ref{Dmenorm})) is now restricted to the unit cell according to
\begin{eqnarray}
&&\frac{1}{\epsilon_0L_{\rm cell}}\int_{\rm cell}\drv\left[1-\real\Gamma(\rv,\om_{\sigma k})+\frac{\om_{\sigma k}}{2}\left.\frac{\partial\real\Gamma(\rv,\om)}{\partial \om}\right\vert_{\om_{\sigma k}}\right]\nonumber\\
&&\times\dv^*_{\sigma k}(\rv)\cdot\dv_{\sigma k}(\rv)\nonumber\\
&=&\frac{\hbar\om_{\sigma k}}{2}.
\label{dmenorm}
\end{eqnarray}
We perform a Taylor expansion of these mode fields about a reference wavevector $\bar{k}$ as
\begin{equation}
\dv_{\sigma k}(\rv)=\sum_{l=0}\frac{(k-\bar{k})^l}{l!}\left.\frac{\partial^l\dv_{\sigma k}(\rv)}{\partial k^l}\right\vert_{k=\bar{k}},
\label{dvTex}
\end{equation}
and define the EP field amplitude through
\begin{equation}
a_{\sigma\bar{k}}(x,t)=\exp\left(\rmi\om_{\sigma\bar{k}}t\right)\intf\dk\langle\aop_{\sigma k}(t)\rangle\exp\left[\rmi(k-\bar{k})x\right],
\label{adef}
\end{equation}
so that we may rewrite (\ref{Dex}) in the form
\begin{eqnarray}
\langle\Dop(\rv)\rangle&\simeq&\sum_\sigma\exp\left[\rmi(\bar{k}x-\om_{\sigma\bar{k}}t)\right]\sum_{l=0}\frac{(-\rmi)^l}{l!}\left.\frac{\partial^l\dv_{\sigma k}(\rv)}{\partial k^l}\right\vert_{k=\bar{k}}\frac{\partial^l a_{\sigma\bar{k}}(x,t)}{\partial x^l}\nonumber\\
&&+c.c..
\label{DTex}
\end{eqnarray}
We observe that to lowest order in $k$, which corresponds to weak mode dispersion and/or narrow-band signals, the EP fields represent spatial envelope functions for each Bloch mode, viz.,
\begin{equation}
\langle\Dop(\rv)\rangle\simeq \sum_\sigma\dv_{\sigma\bar{k}}(\rv)a_{\sigma\bar{k}}(x,t)\exp\left[\rmi (\bar{k}x-\om_{\sigma\bar{k}}t)\right]+c.c..
\label{Dapproxk}
\end{equation}
We then obtain a dynamical equation for the EP fields for $t>0$ from (\ref{aevcs}) and (\ref{adef}) as
\begin{equation}
\frac{{\rm d}a_{\sigma\bar{k}}(x,t)}{{\rm d}t}=-\Delta_{\sigma\bar{k}} a_{\sigma\bar{k}}(x,t)-v^{(\rm g)}_{\sigma\bar{k}}\frac{\partial a_{\sigma\bar{k}}(x,t)}{\partial x},
\label{aeq}
\end{equation}
where the mode frequency $\om_{\sigma k}$ has been approximated by the corresponding Taylor polynomial in $k$ truncated at the first order, and we have defined the Bloch mode group velocity $v^{(\rm g)}_{\sigma\bar{k}}=\partial\om_{\sigma k}/\partial k|_{\bar{k}}$. Working in the regime where the temporal variation of $a_{\sigma\bar{k}}(x,t)$ is such that the time derivative in (\ref{aeq}) may be neglected we then obtain
\begin{equation}
\frac{\partial a_{\sigma\bar{k}}(x,t)}{\partial x}=-\frac{\gamma_{\sigma\bar{k}}}{2}a_{\sigma\bar{k}}(x,t),
\label{aeqss}
\end{equation}
with solutions
\begin{equation}
|a_{\sigma\bar{k}}(x,t)|^2=|a_{\sigma\bar{k}}(0,t)|^2\exp\left(-\gamma_{\sigma\bar{k}} x\right),
\label{asol}
\end{equation}
where the BLB absorption coefficient $\gamma_{\sigma\bar{k}}$ follows from (\ref{Deltadefnorm}) and the normalisation of the Bloch modes (\ref{dmenorm}) as
\begin{equation}
\gamma_{\sigma\bar{k}}=\frac{2\Delta_{\sigma\bar{k}}}{v^{(\rm g)}_{\sigma\bar{k}}}=\frac{2\int_{\rm cell}\drv\imag\Gamma(\rv,\om_{\sigma\bar{k}})\dv^*_{\sigma\bar{k}}(\rv)\cdot\dv_{\sigma\bar{k}}(\rv)}{v^{\rm (g)}_{\sigma\bar{k}}\varepsilon_0\hbar L_{\rm cell}}.
\label{gdef}
\end{equation}
From standard electromagnetic theory we may identify the EP energy dissipated in a unit cell in an optical cycle as
\begin{equation}
E_{\rm diss}=\frac{2\int_{\rm cell}\drv\imag\Gamma(\rv,\om_{\sigma\bar{k}})\dv^*_{\sigma\bar{k}}(\rv)\cdot\dv_{\sigma\bar{k}}(\rv)}{\varepsilon_0L_{\rm cell}},
\label{Edissdef}
\end{equation}
while the EP energy in a unit cell is $E_{\rm phot}=\hbar\om_{\sigma\bar{k}}$. Defining the EP mode quality factor $Q_{\sigma\bar{k}}=E_{\rm phot}/E_{\rm diss}$ then allows us to rewrite (\ref{gdef}) in the form
\begin{equation}
\gamma_{\sigma\bar{k}}=\frac{\om_{\sigma\bar{k}}}{v^{\rm (g)}_{\sigma\bar{k}}Q_{\sigma\bar{k}}}.
\label{gdefQ}
\end{equation}
As expected the absorption coefficient is enhanced by reducing both the group velocity and the quality factor of the EP mode. We emphasise, however, that both structural and material dispersion effects are encapsulated within both the Bloch mode group velocity $v^{\rm (g)}_{\sigma\bar{k}}$ and the EP mode quality factor $Q_{\sigma\bar{k}}$. It is therefore not immediately obvious in general what is the net effect of the material dispersion.

\subsection{The role of structural versus material dispersion on the absorption coefficient}
Towards establishing the role of material dispersion we briefly digress to consider homogeneous media. In this case we have plane wave modes and the group velocity deviates from $c$ purely as a result of material dispersion, viz.,
\begin{equation}
v^{\rm (g)}_{\bar{k}}=\frac{c\left[1-\real\Gamma(\om_{\bar{k}})\right]^{3/2}}{\left[1-\real\Gamma(\om_{\bar{k}})+\frac{\om_{\bar{k}}}{2}\left.\frac{\partial\real\Gamma(\om)}{\partial\om}\right\vert_{\om_{\bar{k}}}\right]}.
\label{vghdef}
\end{equation}
With the various material parameters being independent of $\rv$, (\ref{gdef}) then simplifies to
\begin{equation}
\gamma_{\bar{k}}=\frac{\om_{\bar{k}}}{v^{\rm (p)}_{\bar{k}}}\frac{\imag\Gamma(\om_{\bar{k}})}{\left[1-\real\Gamma(\om_{\bar{k}})\right]},
\label{gdefh}
\end{equation}
where the normalisation equivalent to (\ref{dmenorm}) in a homogeneous medium has been used, and $v^{\rm (p)}_{\bar{k}}=c[1-\real\Gamma(\om_{\bar{k}})]^{1/2}$ is the phase velocity of the medium. The result (\ref{gdefh}) expresses the fact that slow light effects from material dispersion alone (as represented by the medium group velocity) do not lead to enhancement of the absorption coefficient.

%We note that this form of the result is independent of the normalisation of the mode fields. From (\ref{Dmeeq}) we have in a homogeneous medium $k=[1-\real\Gamma(\om)]^{1/2}\om/c$. Introducing the local phase and group velocities corresponding to the material as $v^{(m)}_{\rm p}(\rv,\om)=c\left[1-\real\Gamma(\rv,\om)\right]^{1/2}$ and $v^{(m)}_{\rm g}(\rv,\om)=c[1-\real\Gamma(\rv,\om)]^{3/2}/[1-\real\Gamma(\rv,\om)+\om\partial_\om\real\Gamma(\rv,\om)/2]$, respectively, we may then rewrite (\ref{gdef}) in the form

Returning to structured media involving dispersive materials, we may obtain an expression for $v^{\rm (g)}_{\sigma\bar{k}}$ which highlights the contributions due to the structural dispersion and the material dispersion.
The procedure is to solve the nondispersive problem and then include the dispersion through
perturbation theory.
We proceed by again considering the generalised Hermitian problem (\ref{Deq0h}), which in the periodic geometry here takes the form
\begin{eqnarray}
&&\left[1-\real\Gamma(\rv,\om_{\sigma\bar{k}})\right]\nabv\times\nabv\times\left\{\left[1-\real\Gamma(\rv,\om_{\sigma\bar{k}})\right]\tilde{\Dv}_{\sigma k}(\rv)\right\}\nonumber\\
&&=\frac{\tilde{\om}_{\sigma k}^2}{c^2}\left[1-\real\Gamma(\rv,\om_{\sigma\bar{k}})\right]\tilde{\Dv}_{\sigma k}(\rv).
\label{DmeqHk}
\end{eqnarray}
Recall that the material dispersion has been removed by fixing the response function
$\Gamma(\rv,\om_{\sigma\bar{k}})$ at the reference frequency corresponding to the reference
wavenumber $\bar{k}$.
The tilded quantities thus correspond to the modes of the structured medium at some other wavenumber $k$ in the absence of dispersion, except at the reference point itself where the exact dispersive solution $\Dv_{\sigma\bar{k}}(\rv)$ is by construction a solution to (\ref{DmeqHk}) with $\tilde{\om}_{\sigma\bar{k}}=\om_{\sigma\bar{k}}$. 

Now let us consider a different solution to the full dispersive problem (\ref{Deqm}) with frequency $\om_{\sigma k}$ for $k\simeq \bar{k}$, and define $\delta \tilde{\om}= \om_{\sigma k}- \tilde{\om}_{\sigma k}$ to be the frequency difference between the dispersive and nondispersive solutions at $k$.  On the other hand, the difference in the medium response at the two exact solution frequencies $\om_{\sigma k}$ and $\om_{\sigma\bar{k}}$ is given to first order by $\delta\Gamma(\rv,\om_{\sigma k})=(\om_{\sigma k}-\om_{\sigma\bar{k}})\partial\real\Gamma(\rv,\om)/\partial\om |_{\om_{\sigma\bar{k}}}$.  Then applying standard perturbation theory to the Hermitian problem (\ref{DmeqHk}) at $\om_{\sigma k}$, we may find $\delta \tilde{\om}$ to first order as 
\begin{eqnarray}
\delta\tilde{\om}\simeq-\left(\om_{\sigma k}-\om_{\sigma\bar{k}}\right)\frac{\tilde{\om}\int_{\rm cell}\drv\left.\frac{\partial\real\Gamma(\rv,\om)}{\partial\om}\right\vert_{\om_{\sigma\bar{k}}}\tilde{\Dv}^*_{\sigma k}\cdot\tilde{\Dv}_{\sigma k}}{2\int_{\rm cell}\drv\left[1-\real\Gamma(\rv,\om_{\sigma\bar{k}})\right]\tilde{\Dv}^*_{\sigma k}\cdot\tilde{\Dv}_{\sigma k }}.
\label{ompert}
\end{eqnarray}
This is arbitrarily accurate in the limit as $\delta k=k-\bar{k}\to 0$.  Therefore we may write
\begin{eqnarray}
\lim_{\delta k\to 0}\frac{\delta\tilde\om}{\delta k}&=&\lim_{\delta k\to 0}\frac{\om_{\sigma k}-\tilde\om_{\sigma k}}{\delta k}\nonumber\\
&=&\lim_{\delta k\to 0}\frac{\om_{\sigma k}-\om_{\sigma\bar{k}}}{\delta k}-\lim_{\delta k\to 0}\frac{\tilde\om_{\sigma k}-\om_{\sigma\bar{k}}}{\delta k}\nonumber\\
&=&\lim_{\delta k\to 0}\frac{\om_{\sigma k}-\om_{\sigma\bar{k}}}{\delta k}-\lim_{\delta k\to 0}\frac{\tilde\om_{\sigma k}-\tilde{\om}_{\sigma\bar{k}}}{\delta k}\nonumber\\
&=&v_{\sigma\bar{k}}^{\rm (g)}-\tilde{v}_{\sigma\bar{k}}^{\rm (g)},
\label{limvg}
\end{eqnarray}
Dividing (\ref{ompert}) by $\delta k$, applying (\ref{limvg}) and finally rearranging the result yields
\begin{equation}
v^{\rm (g)}_{\sigma\bar{k}}=\tilde{v}^{\rm (g)}_{\sigma\bar{k}}\frac{\int_{\rm cell}\drv\left[1-\real\Gamma(\rv,\om)\right]\dv^*_{\sigma\bar{k}}(\rv)\cdot\dv_{\sigma\bar{k}}(\rv)}{\int_{\rm cell}\drv\left[1-\real\Gamma(\rv,\om)+\frac{\om}{2}\left.\frac{\partial\real\Gamma(\rv,\om)}{\partial\om}\right\vert_{\om_{\sigma\bar{k}}}\right]\dv^*_{\sigma\bar{k}}(\rv)\cdot\dv_{\sigma\bar{k}}(\rv)},
\label{vgexact}
\end{equation}
where we have exploited the definition of the Bloch modes to replace the
$\lim_{\delta k\to 0}\tilde{\Dv}_{\sigma \bar{k}}(\rv)=\Dv_{\sigma \bar{k}}(\rv)$ with $\dv_{\sigma \bar{k}}$.  This expression shows clearly how the dispersion expressed by $\partial \real \Gamma/\partial \om$ modifies the purely structural group velocity $\tilde{v}^{\rm (g)}_{\sigma \bar{k}}$. 

We can now investigate how the different contributions to the group velocity influence the absorption coefficient.
Substituting for $v^{\rm (g)}_{\sigma\bar{k}}$ in (\ref{gdef}) using (\ref{vgexact}) and the normalisation condition (\ref{dmenorm}) we obtain
\begin{equation}
\gamma_{\sigma\bar{k}}=\frac{\om_{\sigma\bar{k}}}{\tilde{v}^{\rm (g)}_{\sigma\bar{k}}}\frac{\int_{\rm cell}\drv\imag\Gamma(\rv,\om_{\sigma\bar{k}})\dv_{\sigma\bar{k}}^*(\rv)\cdot\dv_{\sigma\bar{k}}(\rv)}{\int_{\rm cell}\drv\left[1-\real\Gamma(\rv,\om_{\sigma\bar{k}})\right]\dv_{\sigma\bar{k}}^*(\rv)\cdot\dv_{\sigma\bar{k}}(\rv)}.
\label{gdefg}
\end{equation}
To express (\ref{gdefg}) in terms of more familiar quantities we identify the electric field Bloch modes from (\ref{Eopa}) as $\bi{e}_{\sigma\bar{k}}(\rv)=\left[1-\real\Gamma(\rv,\om_{\sigma\bar{k}})\right]\dv_{\sigma\bar{k}}(\rv)/\epsilon_0$ and relate $\epsilon(\rv, \om)$ and $\Gamma(\rv,\om)$ to lowest order in the dissipative terms, viz., $\real\,\epsilon(\rv,\om)=1/[1-\real\Gamma(\rv,\om)]$ and $\imag\epsilon(\rv,\om)=\imag\Gamma(\rv,\om)/[1-\real\Gamma(\rv,\om)]^2$. This allows (\ref{gdefg}) to be rewritten as
\begin{equation}
\gamma_{\sigma\bar{k}}=\frac{\om_{\sigma\bar{k}}}{\tilde{v}^{\rm (g)}_{\sigma\bar{k}}}\frac{\int_{\rm cell}\drv\imag\;\epsilon(\rv,\om_{\sigma\bar{k}})\bi{e}_{\sigma\bar{k}}^*(\rv)\cdot\bi{e}_{\sigma\bar{k}}(\rv)}{\int_{\rm cell}\drv\real\;\epsilon(\rv,\om_{\sigma\bar{k}})\bi{e}_{\sigma\bar{k}}^*(\rv)\cdot\bi{e}_{\sigma\bar{k}}(\rv)}.
\label{gdefewd}
\end{equation}
Identifying the quality factor for the mode $\bi{e}_{\sigma\bar{k}}(\rv)$ in a nondispersive medium corresponding to (\ref{DmeqHk}) as
\begin{equation}
\tilde{Q}_{\sigma\bar{k}}=\frac{\int_{\rm cell}\drv\real\;\epsilon(\rv,\om_{\sigma\bar{k}})\bi{e}_{\sigma\bar{k}}^*(\rv)\cdot\bi{e}_{\sigma\bar{k}}(\rv)}{\int_{\rm cell}\drv\imag\;\epsilon(\rv,\om_{\sigma\bar{k}})\bi{e}_{\sigma\bar{k}}^*(\rv)\cdot\bi{e}_{\sigma\bar{k}}(\rv)},
\label{Qnddef}
\end{equation}
we may rewrite (\ref{gdefewd}) in the form
\begin{equation}
\gamma_{\sigma\bar{k}}=\frac{\om_{\sigma\bar{k}}}{\tilde{v}^{\rm (g)}_{\sigma\bar{k}}\tilde{Q}_{\sigma\bar{k}}}.
\label{gdefQ1}
\end{equation}
In contrast to the equivalent expression (\ref{gdefQ}), it is clear that (\ref{gdefQ1}) is completely independent of the variation of $\epsilon(\rv,\om)$ from its value at $\om=\om_{\sigma\bar{k}}$ since it involves only tilded quantities. Therefore, since $\tilde{v}^{\rm (g)}_{\sigma\bar{k}}$ departs from $\tilde{v}^{\rm (p)}_{\sigma\bar{k}}$ purely as a result of structural effects, if one understands the appearance of $1/\tilde{v}^{\rm (g)}_{\sigma\bar{k}}$ in (\ref{gdefQ1}) as identifying the slow light enhancement of the absorption, we can say that material dispersion has no effect upon the slow light enhancement of absorption in a structured medium. Indeed, the expression (\ref{gdefQ1}) is identical to that derived in the complete absence of dispersion \cite{Mortensen:2007}. This represents a generalisation of earlier results concerning translationally invariant, weakly guiding structures where the same lack of enhancement was derived \cite{Mork:2010}. Although the result (\ref{gdefewd}) was derived under the assumption of 1-D periodicity, it may be straightforwardly extended to 2-D and 3-D periodic structures with the same conclusion.
 
\section{Concluding remarks
\label{CR}
}
We have shown that in weakly dissipative media EP operators arise as natural superpositions of the DP operators and lead to a intuitive and convenient description of the absorptive decay of the quantised electromagnetic field. The mode fields associated with the EP excitations account for the structural and material dispersion of the medium and correspond to the results obtained in the non-absorptive regime. We have found that a quite general class of initial states may be constructed from the EP operators while dispensing with the need to consider the state of the medium explicitly, thus reducing enormously the number of degrees of freedom that must be considered in order to specify the evolution of the system uniquely. In addition, the theory presented allows a generalisation of the BLB law for structured media involving lossy, dispersive materials. Our analysis demonstrates that, for light propagation in homogeneous and periodically structured media, material dispersion has no effect upon the BLB absorption coefficient.

\section*{Acknowledgements
}
This research was supported by the Australian Research Council Centre of Excellence for Ultrahigh bandwidth Devices for Optical Systems (Project No. CE110001018). JES is supported by the National Science and Engineering Research Council of Canada (NSERC). ACJ thanks Dr Luke Helt for helpful comments on the manuscript.

\section*{Appendix}
We now present the expressions corresponding to (\ref{Dopex}) for the other electromagnetic field variables. It has been shown \cite{Bhat:2006} that the electric displacement modes $\Dv_{m\om}$ may be used to construct the polarisation field operator as
\begin{equation}
\Pop(\rv)=\sum_m\inth\dom\left[\real\Gamma(\rv,\om)-\frac{z_{m\om}}{\pi}\imag\Gamma(\rv,\om)\right]\Dv_{m\om}(\rv)\cop_{m\om}+H.a..
\label{Popex}
\end{equation}
This expression may be rearranged to give
\begin{eqnarray}
\Pop(\rv)=&\sum_m\inth\dom\Gamma(\rv,\om)\Dv_{m\om}(\rv)\cop_{m\om}+H.a.\nonumber\\
&-\sum_m\inth\dom\left[\frac{z_{m\om}+\rmi\pi}{\pi}\right]\imag\Gamma(\rv,\om)\Dv_{m\om}(\rv)\cop_{m\om}+H.a.,
\label{Popn}
\end{eqnarray}
where the term on the first line on the right hand side of (\ref{Popn}) is a causal relation reminiscent of the corresponding classical expression, while the term on the second line is identified as the noise polarisation.
Using the relation $\Dop(\rv)=\epsilon_0\Eop(\rv)+\Pop(\rv)$ with (\ref{Popex}), the electric field operator then follows as
\begin{eqnarray}
\Eop(\rv)&=&\frac{1}{\epsilon_0}\sum_m\inth\dom\left[1-\real\Gamma(\rv,\om)+\frac{z_{m\om}}{\pi}\imag\Gamma(\rv,\om)\right]\Dv_{m\om}(\rv)\cop_{m\om}\nonumber\\
&&+H.a..
\label{Eopex}
\end{eqnarray}
Finally, the magnetic induction operator is constructed as
\begin{equation}
\Bop(\rv)=\sum_m\inth\dom\Bv_{m\om}(\rv)\cop_{m\om}+H.a.,
\label{Bopex}
\end{equation}
with the magnetic induction modes obtained from
\begin{equation}
\Bv_{m\om}(\rv)=-\frac{\rmi}{\epsilon_0\om}\nabv\times\left\{\left[1-\real\Gamma(\rv,\om)+\frac{z_{m\om}}{\pi}\imag\Gamma(\rv,\om)\right]\Dv_{m\om}(\rv)\right\},
\label{Bmdef}
\end{equation}
and the magnetic field operator is given by $\Hop(\rv)=\Bop(\rv)/\mu_0$. Alternatively, (\ref{Dmeq}) and (\ref{Bmdef}) may be combined to obtain an equation for the magnetic induction modes,
\begin{equation}
\nabv\times\left\{\left[1-\real\Gamma(\rv,\om)+\frac{z_{m\om}}{\pi}\imag\Gamma(\rv,\om)\right]\nabv\times\Bv_{m\om}(\rv)\right\}=\frac{\om^2}{c^2}\Bv_{m\om}(\rv),
\label{Bmeq}
\end{equation}
which are then normalised according to
\begin{equation}
\frac{c^2}{\om^2}\frac{\pi}{\mu_0\hbar}\int\drv\imag\Gamma(\rv,\om)\left[\nabv\times\Bv^*_{m\om}(\rv)\right]\cdot\left[\nabv\times\Bv_{m\om}(\rv)\right]=\frac{\pi^2}{z_{m\om}^2+\pi^2},
\label{Bmnorm}
\end{equation}
and the electric displacement modes then follow from
\begin{equation}
\Dv_{m\om}(\rv)=\frac{\rmi}{\mu_0\om}\nabv\times\Bv_{m\om}(\rv).
\label{Dmdef}
\end{equation}


\section*{References}
\begin{thebibliography}{10}
\bibitem{Lax:1971} Lax M and Nelson D F 1971 {\it Phys. Rev. B} {\bf 4} 3694
\bibitem{Drummond:1990} Drummond P D 1990 {\it Phys. Rev. A} {\bf 42} 6845
\bibitem{Glauber:1991} Glauber R J and Lewenstein M 1991 {\it Phys. Rev. A} {\bf 43} 467
\bibitem{Nelson:1994} Nelson D F and Chen B 1994 {\it Phys. Rev. B} {\bf 50} 1023
\bibitem{Matloob:1995} Matloob R, Loudon R, Barnett S M and Jeffers J 1995 {\it Phys. Rev. A} {\bf 52} 4823
\bibitem{Gruner:1996} Gruner T and Welsch D G 1996 {\it Phys. Rev. A} {\bf 53} 1818
\bibitem{Dung:1998} Dung H T, Kn{\"o}ll L and Welsch D G 1998 {\it Phys. Rev. A} {\bf 57} 3931
\bibitem{Dung:2003} Dung H T, Buhmann S Y, Kn{\"o}ll L, Welsch D G, Scheel S and K{\"a}stel J 2003 {\it Phys. Rev. A} {\bf 68} 043816
\bibitem{Suttorp:2004} Suttorp L G and Wubs M 2004 {\it Phys. Rev. A} {\bf 70} 013816
\bibitem{Bhat:2006} Bhat N A R and Sipe J E 2006 {\it Phys. Rev. A} {\bf 73} 063808
\bibitem{Suttorp:2007} Suttorp L G 2007 {\it J. Phys. A: Math. Theor.} {\bf 40} 3697
\bibitem{Raabe:2007} Raabe C, Scheel S and Welsch D G 2007 {\it Phys. Rev. A} {\bf 75} 053813
\bibitem{Philbin:2010} Philbin T G 2010 {\it New J. Phys.} {\bf 12} 123008
\bibitem{Judge:2013} Judge A C, Steel M J, Sipe J E and de Sterke C M 2013 {\it Phys. Rev. A} {\bf 87} 033824
\bibitem{Jauch:1948} Jauch J M and K M Watson 1948 {\it Phys. Rev.} {\bf 74} 950
\bibitem{Huttner:1992} Huttner B and Barnett S M 1992 {\it Phys. Rev. A} {\bf 46} 4306
\bibitem{Philbin:2011} Philbin T G 2011 {\it New J. Phys.} {\bf 13} 063026
\bibitem{Yao:2009} Yao P, Van Vlack C, Reza A, Patterson M and Dignam M M 2009 {\it Phys. Rev. B} {\bf 80} 195106
\bibitem{Hopfield:1958} Hopfield J J 1958 {\it Phys. Rev.} {\bf 112} 1555
\bibitem{Barnett:1997} Barnett S M and Radmore P M 1997 {\it Methods in Theoretical Quantum Optics} (Oxford: Oxford University Press)
\bibitem{Walls:1994} Walls D F and Milburn G J 1994 {\it Quantum Optics } (Berlin: Springer-Verlag)
\bibitem{Louisell:1973} Louisell W H 1973 {\it Quantum Statistical Properties of Radiation } (New York: Wiley)

\bibitem{Yang:2008} Yang Z, Liscidini M and Sipe J E 2008 {\it Phys. Rev. A} {\bf 77} 033808
\bibitem{Berestetski:1984} Berestetsk{\u\i} V B, Lifshitz E M and Pitaevski{\u\i} L P 1984 {\it Quantum Electrodynamics} (Oxford: Pergamon)
\bibitem{Khurgin:2009} Khurgin J. B. and Tucker R S 2009 {\it Slow Light: Science and Applications} (Boca Raton: CRC Press, Taylor \& Francis Group)
\bibitem{Mortensen:2007} Mortensen N A and Xiao S 2007 {\it Appl. Phys. Lett} {\bf 90} 141108
\bibitem{Mork:2010} M{\o}rk J and Nielsen T R 2010 {\it Opt. Lett.} {\bf 35} 2834
\bibitem{Fano:1961} Fano U 1961 {\it Phys. Rev.} {\bf 124} 1866
\bibitem{Dirac:1927} Dirac P 1927 {\it Z. Phys. A} {\bf 44} 585
\bibitem{Landau:1960} Landau L D and Lifshitz E M 1960 {\it Electrodynamics of Continuous Media}
(Oxford: Pergamon Press)
\bibitem{Chin:2009} Chin S, Decaire I, Beugnot J C, Mafang S F, Herraez M G and Th{\'e}venaz L 2009 {\it Topical Meeting on Slow and Fast Light, OSA Technical Digest} (Optical Society of America) paper SMA3
\bibitem{Power:1959} Power E A and Zienau S 1959 {\it Philos. Trans. R. Soc. London A} {\bf 251} 427
\bibitem{Merzbacher:1970} Merzbacher, E 1970 {\it Quantum Mechanics} (New York: John Wiley \& Sons)
\bibitem{Sipe:2009} Sipe, J E 2009 {\it J. Opt. A: Pure Appl. Opt.} {\bf 11} 114006
\end{thebibliography}
\end{document}